\documentclass[10pt,letterpaper]{article}
\usepackage{opex3}

\begin{document}

\title{Sub-diffraction thin-film sensing with planar terahertz metamaterials}

\author{Withawat Withayachumnankul$^{1,2}$, Hungyen Lin$^1$, Kazunori Serita$^3$, Charan M. Shah$^4$, Sharath Sriram$^4$, Madhu Bhaskaran$^4$,\\Masayoshi Tonouchi$^3$, Christophe Fumeaux$^1$, and Derek Abbott$^1$}

\address{$^1$School of Electrical \& Electronic Engineering, \\The University of Adelaide, Adelaide, SA 5005, Australia\\

$^2$Faculty of Engineering, \\King Mongkut's Institute of Technology Ladkrabang, Bangkok 10240, Thailand\\

$^3$Institute of Laser Engineering, \\Osaka University, 2-6 Yamadaoka, Suita, Osaka 565-0871, Japan\\

$^4$Functional Materials and Microsystems Research Group, \\
School of Electrical and Computer Engineering, \\RMIT University, Melbourne, VIC 3001, Australia}

\email{withawat@eleceng.adelaide.edu.au} 

\begin{abstract}
Planar metamaterials have been recently proposed for thin dielectric film sensing in the terahertz frequency range. Although the thickness of the dielectric film can be very small compared with the wavelength, the required area of sensed material is still determined by the diffraction-limited spot size of the terahertz beam excitation. In this article, terahertz near-field sensing is utilized to reduce the spot size. By positioning the metamaterial sensing platform close to the sub-diffraction terahertz source, the number of excited resonators, and hence minimal film area, are significantly reduced. As an additional advantage, a reduction in the number of excited resonators decreases the inter-cell coupling strength, and consequently the resonance Q factor is remarkably increased. The experimental results show that the resonance Q factor is improved by 113\%. Moreover, for a film with a thickness of $\lambda/375$ the minimal area can be as small as $0.2\lambda\times0.2\lambda$. The success of this work provides a platform for future metamaterial-based sensors for biomolecular detection.
\end{abstract}



\section{Introduction}

Many significant microscopic mechanisms in materials have strong resonant oscillations at terahertz frequencies. These include, e.g., plasma oscillation, damping in moderately doped semiconductors \cite{Exter90} and intermolecular motions in biomolecules \cite{Fischer07}. Conventional terahertz time-domain spectroscopy (THz-TDS) therefore becomes a valuable tool in assessing the static and dynamic functions of these substances. However, a number of substances are present or can be produced only in the form of thin films. This condition implies a short interaction length between the interrogating terahertz wave and the sample, and consequently, a small change in the measured terahertz signal \cite{Withawat08a}. In any realistic system, this small change can be easily overwhelmed by measurement uncertainty \cite{Withawat08b}, which ultimately limits the system sensitivity \cite{Withawat11}. As a solution for thin-film sensing, multiple approaches that augment conventional THz-TDS have been established using different concepts \cite{Ohara12}.

Planar metamaterials or metasurfaces are among several approaches utilized for thin-film sensing at the terahertz frequency range \cite{Driscoll07}. A metamaterial is typically made of a periodic array of subwavelength metallic resonators that are collectively coupled to the free space excitation \cite{Withawat09,Khodasevych12}. The array's effective response is determined from the geometry and dimensions of the resonators, together with the mutual coupling between them. On resonance, each resonator develops a strong electric field typically confined in a small capacitive gap region. When the resonator is covered by a loading thin film, the effective permittivity at the gap is increased. In effect, the gap undergoes a significant change in the charge distribution and capacitance, which can be indirectly observed via a frequency shift in the transmission resonance. This metamaterial-based approach has been successfully applied to a range of thin-films with various designs and configurations \cite{Debus07,Ohara08,Singh11,Tao11,Jansen11}.

Nevertheless, observing the resonance by using a far-field terahertz excitation imposes a limitation on the minimal area of a thin film deposited on a metamaterial. Fundamentally restricted by diffraction, the smallest possible beam waist in free space is approximately $\lambda/2$. In practice the beam waist is usually even larger due to imperfection in optical components, aberrations, and limitation in alignment accuracy. This large beam spot necessitates a relatively large volume of substance to cover the full area of excited resonators. Even worse, exciting a large number of densely coupled resonators causes resonance broadening due to enhanced scattering \cite{Sersic09}. This results in a decreased Q factor, which is detrimental to the sensitivity of a resonance-based sensor. Although possible, employing a transmission line for resonator excitation similar to that carried out at the microwave band \cite{Withawat12} raises the issues of terminal coupling and propagation loss.

This article proposes a combination of metamaterial-based thin-film sensing with a focused-beam terahertz near-field technique. Such a contactless near-field approach is essential for this investigation and is advantageous in terms of simplicity \cite{Withawat07}. In principle, a terahertz beam with a diameter far below $\lambda/2$ can be observed in the near-field region of a nonlinear crystal used for electro-optical terahertz generation. By introducing a metamaterial into this sub-diffraction terahertz beam, only a small number of constituent resonators are activated. The implications are a reduction in the required sample volume and an increase in the sensitivity through an improved Q factor. The concept is validated through experiment with support from rigorous numerical modelling of the sub-wavelength terahertz source. 

\section{Metamaterial design and fabrication}

\begin{figure}[t]
\centering
\includegraphics{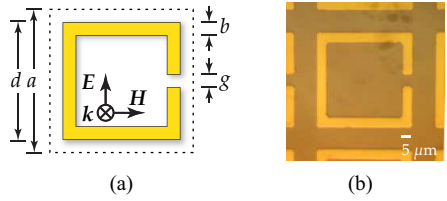}
\caption{Split-ring resonators. (a) Schematic of a single unit cell. The polarization for operation is indicated inside the split ring. The dimensions are as follows: $a=55$ $\mu$m, $d=45$ $\mu$m, $b=5$ $\mu$m, and $g=5$ $\mu$m. (b) Partial view of the fabricated metamaterial under an optical microscope. }\label{NF_schematic}
\end{figure}

A split-ring resonator (SRR), typically used to achieve negative permeability \cite{Pendry99}, is chosen as a primitive for a metamaterial-based thin-film sensor. As shown in Fig.~\ref{NF_schematic}(a), a single SRR is composed of an inductive ring with a capacitive gap. An electric field with polarization perpendicular to the gap excites an oscillating current in the loop and an accumulation of charges at the gap. The resonance occurs when the magnetic energy stored around the loop is equal to the electric energy stored at the gap. As discussed earlier, this gap region is highly sensitive to a change in the surrounding dielectric permittivity. Depositing an analyte onto the gap causes a shift in the resonance. The dimensions of a SRR for this work are specified in Fig.~\ref{NF_schematic}(a). These parameters are selected so that the metamaterial composed of these resonators exhibits a first-order resonance, so-called \emph{LC} resonance, at approximately 0.4~THz.

The metamaterial comprising a periodic array of such SRRs has been defined by microfabrication techniques. A float-zone intrinsic silicon substrate with a resistivity greater than 20,000~$\Omega\cdot$cm and thickness of 380~$\mu$m was cleaned with acetone and isopropyl alcohol. Photolithography with a chlorobenzene dip was performed to define an inverse pattern through a lift-off profile. The substrate was subsequently coated with a film of metal by electron beam evaporation after pumping down to a base pressure of 1$\times$10$^{-7}$~Torr. A 20~nm chromium adhesion layer was deposited followed by a 200~nm thick layer of gold. Lift-off was performed with ultrasonic agitation in acetone and isopropyl alcohol. The wafer defined with the periodic SRR array was sectioned into 10$\times$10 mm$^2$ dies with a rotary diamond dicing saw. An example of the resulting SRR is shown in Fig.~\ref{NF_schematic}(b). 

\section{Terahertz near-field system and numerical model}

The terahertz near-field system employed in this work is based on the focused-beam technique. A fiber-coupled ultrafast laser, operating at a center wavelength of 1.56~$\mu$m, emits the pump beam that is modulated and then focused onto the 200~$\mu$m thick nonlinear crystal, DAST \cite{Schneider06}. The terahertz beam is generated through the optical rectification process. To reduce the spot size of the terahertz beam, the optical beam is first expanded by means of two convex lenses, then focused onto the nonlinear crystal with a beam waist of 25~$\mu$m. Photoconductive sampling, which is used for far-field detection of the terahertz wave, is gated by the optical beam with a center wavelength of 780~nm. Details of the terahertz near-field system employed in this investigation can be found elsewhere \cite{Serita10}.

\begin{figure}[t]
\centering
\includegraphics{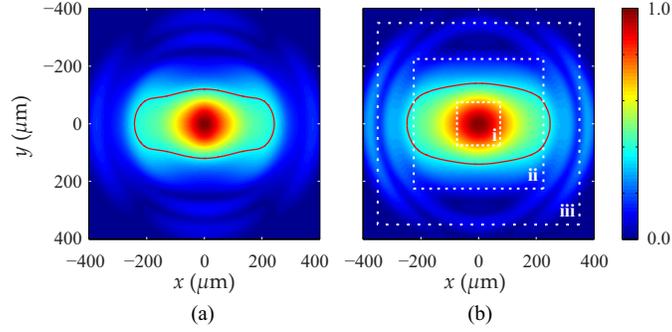}
\caption{Modelled cross sections of terahertz beam, showing the normalized $E_y$ component (a) at a distance of 25~$\mu$m, and (b) at a distance of 50~$\mu$m from the crystal surface. The far-field electric-field polarization is in the y-axis. The cross sections are registered at 0.4~THz. The red lines define the $1/e$ spot size. The rectangular boxes in (b) illustrate the area of test samples from (i) 150$\times$150~$\mu$m$^2$, (ii) 450$\times$450~$\mu$m$^2$, to (iii) 700$\times$700~$\mu$m$^2$.}\label{NF_model}
\end{figure}

The electric-field distribution of the near-field terahertz beam at a given frequency and distance from the crystal surface has been computed through full-wave electromagnetic simulations employing Gaussian aperture sources, which were implemented based on the techniques presented in \cite{Lin10,Lin11}. The parameters of DAST at terahertz and infrared frequency ranges are obtained from \cite{Walther00,Pan96}, respectively. The details of the simulations and its thorough validation will be presented elsewhere. An illustration of the beam profiles at 0.4~THz at a distance of 25 and 50~$\mu$m from the output surface of the DAST crystal is displayed in Fig.~\ref{NF_model}. Only the $E_y$ component, which excites the metamaterial in the sensing process, is shown. The spot sizes at the two distances illustrate the rapid widening of the beam with distance from the sub-wavelength electro-optical source as a result of diffraction.

\section{Experiment}

\begin{figure}[b]
\centering
\includegraphics{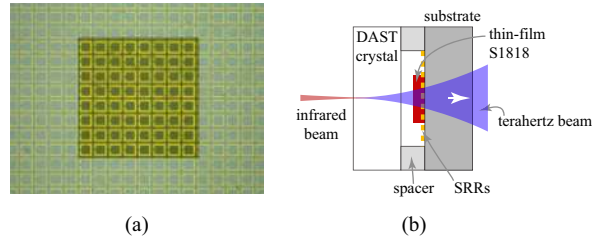}
\caption{Sample and its placement in near-field. (a) Metamaterial coated with photoresist S1818G with an area of 450$\times$450~$\mu$m$^2$ and a thickness of 2~$\mu$m. (b) Schematic (not to scale) showing the location of the sample under measurement by terahertz near-field system.}\label{NF_experiment}
\end{figure}

Photoresist S1818G with $n-j\kappa=1.6-0.02j$ at 0.4~THz is selected as an analyte, owing to its readily controllable patterning through photolithography. To evaluate the performance of the proposed near-field detection scheme, the thickness of the fabricated photoresist film is chosen to be 2~$\mu$m, and the area of the film varies from 150$\times$150, 450$\times$450, 700$\times$700~$\mu$m$^2$, to full plane. A comparison between these film areas and the near-field spot size can be seen in Fig.~\ref{NF_model}(b). Noticeably, the 150$\times$150~$\mu$m$^2$ film is much smaller than the $1/e$ cross section of the beam, whilst the 300$\times$300~$\mu$m$^2$ film is comparable in size to the cross section. In terms of the energy, the 150$\times$150, 450$\times$450, and 700$\times$700~$\mu$m$^2$ samples are exposed to about 30, 84, and 97\% of the total beam energy, respectively.

Fig.~\ref{NF_experiment}(a) shows the sample film deposited on the fabricated metamaterial. Depicted in Fig.~\ref{NF_experiment}(b), the coated metamaterial is placed in the near-field at a distance of approximately 50~$\mu$m from the crystal surface. The sub-diffraction terahertz beam generated from the nonlinear crystal interacts with only a fraction of the SRRs on the metamaterial. Afterwards, the propagating wave is temporally resolved in the far-field. In order to remove the fringe effect caused by reflections in the substrate, the time-resolved waveform is cropped and zero-padded prior to Fourier transform. The sample transmission is then normalized by the reference transmission measured from a bare silicon substrate under the same condition. For the sake of comparison, a conventional far-field THz-TDS system is used to measure the identical metamaterial that is placed at the focal plane. 

\section{Results \& discussion}

\begin{figure}
\centering
\includegraphics{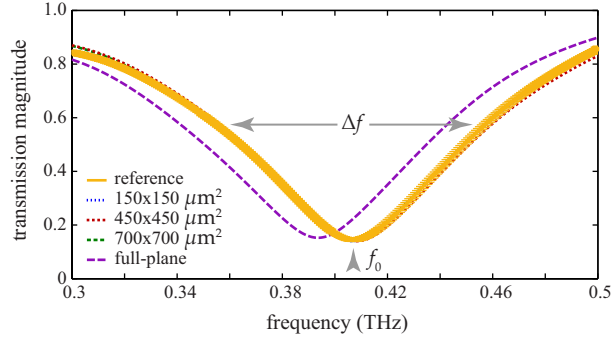}
\caption{Far-field measurement results. The transmission of the uncoated metamaterial is accompanied by the error bars with one standard deviation. The transmission profiles for the metamaterial with photoresist films with an area of 700$\times$700, 450$\times$450, and 150$\times$150~$\mu$m$^2$ are mostly masked by the error bars. The Q factor is calculated from $f_0/\Delta f$.}\label{NF_farfield}
\end{figure}

The results from the far-field measurement are shown in Fig. \ref{NF_farfield}. In order to characterize the system uncertainty, the measurement of the uncoated metamaterial is repeated five times. The transmission of the uncoated metamaterial reveals its resonance frequency at 0.407~THz with a Q factor of 4.0. On resonance, the effective extinction cross section of the uncoated sensor, defined as $a^2(1-T)$, is 2586~$\mu$m$^2$. This result is in accordance with numerical simulation (not shown here). For the full-plane photoresist film, the resonance shifts to 0.394~THz, or equivalently -3.2\% from the base frequency, with a small additional damping due to the sample absorption. This can be converted to the sensitivty of 8.1~GHz/RIU (RIU denotes refractive index unit). For the 700$\times$700 $\mu$m$^2$ film and smaller, a change in the transmission profile compared to the reference is not distinguishable from the measurement uncertainty. 

\begin{figure}
\centering
\includegraphics{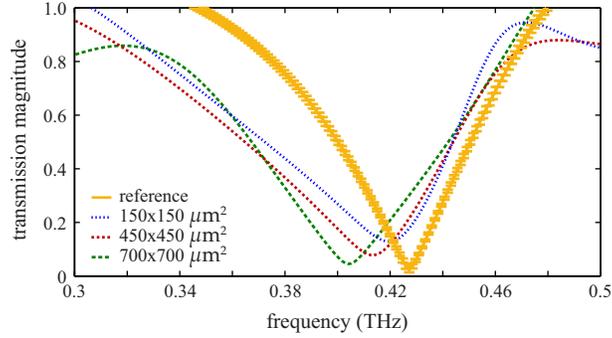}
\caption{Near-field measurement results. The transmission of the uncoated metamaterial is accompanied by the error bars with one standard deviation. The resonance frequency of the uncoated metamaterial is 0.427~THz, and the resonance frequencies in the case of 700$\times$700, 450$\times$450, 150$\times$150~$\mu$m$^2$ films are 0.404, 0.414, and 0.420~THz, respectively. }\label{NF_nearfield}
\end{figure}

Fig. \ref{NF_nearfield} shows the transmission results from the near-field measurement. The uncoated metamaterial, measured five times successively, exhibits its resonance centered at 0.427~THz with a Q factor of 8.54 and an extinction cross section of 2946~$\mu$m$^2$. Compared with the resonance of the identical sample measured in the far-field, the near-field resonance undergoes 5\% blueshift, 113\% increase in the Q factor, and 14\% increase in the effective extinction cross section. This drastic change in the resonance properties can be ascribed to a reduction in the interactions between the resonators. In the far-field measurement, the large beam spot excites a large number of resonators, resulting in reinforced inter-cell coupling. By contrast, the sub-diffraction beam spot in the near-field excites a much smaller number of the resonators. As a result, intercell-couplings become relatively weak. Although not exactly identical, this situation is similar to the resonance change following a modification in the lattice density \cite{Sersic09}. As the density increases, the inter-resonator coupling becomes stronger, leading to a broadened resonance and a reduced extinction cross section. A quantitative analysis of the metamaterial response dependency on the excitation pattern is out of the scope of this work and being investigated separately. 

\begin{figure}
\centering
\includegraphics{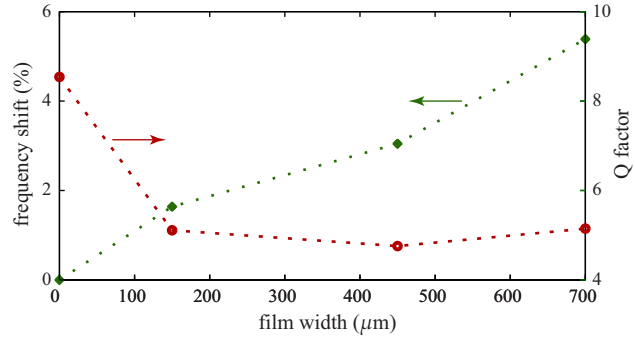}
\caption{Near-field resonance characteristics. The frequency shift and Q factor are plotted as a function of the film width. The frequency shift is with respect to the resonance frequency of the uncoated metamaterial (zero film width). The dotted lines are merely for visual guidance.}\label{NF_characteristic}
\end{figure}

In Fig.~\ref{NF_nearfield}, a change in the resonance in comparison to the reference can be clearly observed for all of the samples. The analyte as small as 150$\times$150~$\mu$m$^2$, equivalent to 3$\times$3 resonators or $0.2\lambda \times 0.2\lambda$ at 0.4~THz, can be unambiguously detected as a clear measurable redshift in the resonance. It is worth noting that the irregularities of the transmission amplitude below 0.35 THz and above 0.48~THz are due to early truncation and zero padding of the time-domain signals to remove the unwanted reflections. This problem can be easily resolved by using a thicker substrate for supporting the resonators. Fig. \ref{NF_characteristic} elucidates the resonance characteristics as a function of the sample size. As the film area becomes larger, the resonance frequency redshifts almost in a linear fashion, whilst the Q factor initially decreases and slightly increases for the 700$\times$700~$\mu$m$^2$ film. 

The resonance behaviour under partial dielectric loading can be explained by the interaction between two competing resonator groups, one with and the other without the loading sample. In the case where the sample area is smaller than the beam spot, both groups of the resonators are excited. The analyte-loaded resonators clustering in the beam center locally resonate at a lower frequency, whilst the unloaded resonators away from the center resonate at a higher frequency. The two local resonances contribute to the far-field resonance broadening, or the lowered Q factor, as is observed with the 150$\times$150 and 450$\times$450~$\mu$m$^2$ samples. This effect is analogous to inhomogeneous broadening in molecular and atomic transitions. It is expected that by removing the uncovered resonators, the Q factor will be restored. As discussed earlier, the 700$\times$700~$\mu$m$^2$ film covers most of the excited resonators. Hence, the observed resonance is homogeneous, resulting in the increased Q factor.

\section{Conclusion}

This article presents enhanced-sensitivity thin-film detection utilizing a planar metamaterial in conjunction with a sub-diffraction terahertz source. Exposing the metamaterial to the near-field source results in a significant decrease in the number of excited resonators. The benefits are twofold: a decrease in the sample volume and an increase in the resonance Q factor. Since only a small number of resonators are activated, the film area required to cover the resonators can be reduced. At the same time, the Q factor of the metamaterial is improved because of weak near-field interactions among a small group of the resonators. The experiment shows that the system can unambiguously sense the $\lambda/375$-thick dielectric film with an area of 3$\times$3 resonators or $0.2\lambda\times 0.2\lambda$. This level of sensitivity is not achievable with a far-field probing technique. In addition, the uncoated metamaterial measured in the near-field exhibits the fundamental resonance with a Q factor of 8.54, which is 113\% improvement in comparison to far-field probing.

Further improvement in the sensitivity is possible on the basis of this proposed near-field sensing scheme. The number of excited resonators can be additionally reduced by bringing the metamaterial-based sensor closer to the near-field source or, even better, by fabricating sub-wavelength resonators directly onto the nonlinear crystal. Another possibility is to substitute SRRs by asymmetric resonators that essentially possess a sharp Fano resonance owing to the dark mode \cite{Debus07,Singh11}. Other sensitivity improvement techniques such as increasing the metal thickness \cite{Chiam09} or reducing the permittivity of the substrate \cite{Tao10} can also be realized under this near-field sensing approach.

\section*{Acknowledgement}

WW, SS, and MB acknowledge Australian Post-Doctoral Fellowships from the Australian Research Council (ARC) through Discovery Projects DP1095151, DP110100262, and DP1092717, respectively. WW acknowledges a travel grant from Australian Nanotechnology Network. HL acknowledges the Prime Minister’s Australia Asia Endeavour Award for research collaboration with Japan. CF acknowledges the ARC Future Fellowship funding scheme under FT100100585. MT acknowledges  fundings by Grant-in-Aid A22246043,  JSPS, and Industry-Academia Collaborative R\&D, JST. DA acknowledges funding via the Australian Research Council (ARC) Discovery Projects DP0881445 and DP1097281.

\end{document}